# Study of the physical properties of selected active objects in the main belt and surrounding regions by broadband photometry

Serhii Borysenko[1;*], Alexander Baransky[2], Ekkehard Kuehrt[3], Stephan Hellmich[3], Stefano Mottola[3], Karen Meech[4]

[1] Main Astronomical Observatory of NAS of Ukraine, Akademika Zabolotnoho Str. 27, 03143 Kyiv, Ukraine

[2] Astronomical Observatory of Taras Shevchenko National University of Kyiv, Observatorna Str. 3, Kyiv 04053, Ukraine

[3] DLR-German Aerospace Center, Institute of Planetary Research, Rutherfordstr. 2, 12489 Berlin, Germany

[4] Institute for Astronomy, University of Hawaii, 2680 Woodlawn Drive, Honolulu, HI 96822, USA

Dynamically different groups of comets and active asteroids with orbits at 2 – 5 a. u. show dust activity in varying degrees and forms. Photometric study and comparison of physical parameters can help to classify mechanisms and nature of the activity for such objects. We present new observations using broadband photometry of 15 active objects in the main asteroid belt and surrounding regions obtained during 2012 – 2016. The study aims to compare the physical properties of main-belt comets (MBCs), quasi-Hilda comets (QHCs), and active asteroids (AAs). The observations were carried out with the 0.7-m telescope of the Kyiv Comet Station, Ukraine, and with the 1.23-m telescope at the Calar Alto Observatory, Spain, using BVR broadband filters. Upper limits for the nuclear radii, *Afp* parameters, and color indices were measured. The results of 2012 – 2016 observations suggest that there exist systematic differences in the physical parameters of MBCs and QHCs.

## 1 Introduction

Over the past decade, several dozen objects have been discovered in nearly-circular orbits within the main asteroid belt (MB), which periodically or occasionally display an ex-tended appearance in the form of a coma and/or a tail that is generally typical for comets. These objects are collectively referred to as Main Belt Active Objects or MBAOs (Hsieh 2017). Such objects have various morphology of comae and tails as a result of the associated activity of different nature (sublimation, impact, rotational effects, or other). According to Hsieh & Jewitt (2006), no known mechanism that could drive bodies from the outer region of the Solar System into near-circular orbits in the main asteroid belt. Hence, MBAOs are likely the bodies that were formed in the inner part of the Solar System and genetically related to asteroids. Based on the type of their activity, these objects are generally classified as Active Asteroids (AAs) if they display just a dust activity, and Main Belt Comets (MBCs) if they additionally show evidence of a sublimation-driven activity.

The fact that some native objects from the MB can show gas release reinforces the notion that many asteroids – especially those located in the outer part of the MB – are rich in volatiles that are not usually observed because they would be buried under a desiccated, inert layer (Jewitt et al. 2015b). MBCs are, therefore, particularly interesting because they provide an opportunity to study materials that could be hidden in the interior of MB asteroids and help us understand the formation mechanisms and composition of the MB (Jewitt 2012; Jewitt et al. 2015c). Further, the release of volatiles from MBCs bears interest also because impacting volatile-rich asteroids may represent a better mechanism for delivering water to the early Earth than impacts of classical comets (Morbidelli et al. 2000) since the latter doesn't seem to have a compatible deuterium-hydrogen ratio.

The apparent brightness of MBAOs depends on the strength of their activity, nucleus size, heliocentric and geo-centric ranges. It also depends on the dust grain properties, such as grain size and composition (albedo). The morphology of MBAOs is diverse: they can show comae, and a long, thin, broad, or curved tail. Some objects can even have multiple tails. Some are active for a long time, while others are active only during a few weeks (Hsieh & Jewitt 2006). Objects with repeated activity on multiple apparitions are probably driven by sublimation, while, some one-time events are consistent with collisions.

While the mechanisms responsible for the mass loss in MBCs and collisional events are clear, they are not understood for the majority of AAs (Hirabayashi et al. 2014). Possible causes are rotational mass shedding and thermally-induced stress release (Hsieh et al. 2018, Lauretta et al. 2019).

---

[*] Corresponding author: borisenk@mao.kiev.ua

In addition to MBAOs, also Quasi-Hilda comets (QHCs) are of particular interest. Those are Jupiter-family comets (JFCs) that are associated with the Hilda asteroid zone. However, in contrast to Hilda asteroids, which reside in a stable 3:2 mean motion resonance with Jupiter, quasi-Hilda objects are on unstable orbits (Toth 2006). It was shown that Centaurs and transneptunian objects could be injected into the quasi-Hilda region (Gil-Hutton & Garcıa-Migani 2016). Di Sisto et al. (2005) found that asteroids escaped from the stable 3:2 Hilda population can also end up as quasi-Hilda objects. Some QHCs (Lowry & Fitzsimmons 2001) can display dust tails similar to MBAOs (Borysenko et al. 2019; Shi et al. 2019), which indicates a common nature of the activity for these objects. Therefore, it is useful to compare dust activity for both groups.

## 2 Observational methods

A cometary spectrum can contain two distinct components: 1) a continuum scattered by the central nucleus and coma dust grains, and 2) an emission spectrum from the coma and tail excited gases. Usually, narrowband filters centred at specific wavelengths are used to obtain quantitative data for both components (Schleicher & Farnham 2004; A'Hearn & Cowan 1975). Narrowband images are needed especially for the estimation of gas production rates. However, since QHCs and MBAOs have a total apparent magnitude generally fainter than V=18 at opposition, they are accessible from small telescopes (d < 2m) only through broadband observations.

Broadband filters (*BVRI*) were introduced for stellar photometry and were not designed for observing comets. However, those filters can be useful for general cometary observations of distant or faint objects, or objects that mainly display a dust component, as they sample the reflected solar continuum well and have a large through-put. Broadband observations over time are also valuable to monitor the object's time-variable behaviour (Swamy 2010; Bauer et al. 2012).

Observations in the red band can be used for computing the *Afρ* quantity as a proxy for the amount of dust in MBAOs and QHCs (Meech & Svoren 2004, A'Hearn et al. 1984). Although deriving an absolute dust production rate from *Afρ* requires an assumption about the (unknown) dust size distribution, the use of *Afρ* is still very useful for comparing the activity level of an object at different epochs, or the activity levels of different objects. With the additional use of blue and visual filters, we can obtain color indices, which could help us to resolve subgroups of comets and to obtain information about the evolution or the age of these objects (Jewitt 2015). However, more detailed and systematic information is a need in general. Our work gives only some part of such data that can be used for future analysis.

## 3 Observations

In the period 2012 – 2016, observations of selected targets were taken by Alexander Baransky and Serhii Borysenko at the Kyiv Comet Station (MPC code – 585), Lisnyky, Ukraine. The standard Johnson broadband filters *R* and *V* were used with an FLI PL 47-10 CCD camera at the prime focus (F = 2828 mm) of the 0.7 – m reflector. The detector consists of a 1024 × 1024 array of 13 μm pixels, which corresponds to a scale of 0.947″ per pixel. The typical read-out noise is about 10 $\bar{e}$ and the conversion gain is 1.22 $\bar{e}$/ADU.

During the same period, R filtered observations of target objects were made at the Calar Alto (CA) Observatory, Spain, with the 1.23-m telescope (MPC code – G36) and an e2v CCD231-84-NIMO-BI-DD CCD sensor with 4k × 4k 15 μm pixels. The camera is installed at the telescope's Ritchey-Chretien focus (F = 9857 mm) with no field corrector and has a plate scale of 0.314″/pixel. Binning 2×2 was used, resulting in an effective pixel scale of 0.628″.

## 4 Analysis

Image processing was done with the Astroart 4.0 software (http://www.msb-astroart.com/), while stacking the frames was done using Astrometrica 4.0 (http://www.astrometrica.at/). We used ATV IDL (Barth et al. 2001) routines for aperture photometry of comets and reference stars. Aperture radii of about 2.5″– 7.5″ were used for Calar Alto observations and about 7.6″– 11.4″ for Kyiv Comet Station observations, except in the cases of close star appulses, for which the aperture size was reduced.

The APASS-9 star catalog was used as a photometric reference (Henden et al. 2016). This catalog includes magnitudes of stars from about 7[th] magnitude to about 17[th] magnitude in five filters: Johnson *B* and *V*, plus Sloan g', r', i'. It has mean uncertainties of about 0.07 mag for *B*, about 0.05 mag for *V*, and less than 0.03 mag for r'. We used the transformation formula from r' to $R_C$ magnitudes derived by Munari et al. (2014). About 8 – 12 reference stars of 14[th] – 17[th] magnitude were used for each night. The main source of photometric errors was the signal-to-noise ratio (*SNR*) of the target. Generally, in the *B* filter the *SNR* was about 5 – 10, in the *V* filter 12 – 20, and more than 20 in *R*.

The *Afρ* parameter was calculated from (A'Hearn et al. 1984):

$$Af\rho = 4r^2\Delta^2 \cdot 10^{0.4(m_\odot - m_c)}\rho^{-1}, \qquad (1)$$

here r (AU) is the heliocentric distance; Δ (cm) is the geocentric distance; $m_\odot$, $m_c$ are the apparent *R* magnitudes of the Sun and of the comet, respectively; ρ [cm] is the radius of the photometric aperture projected onto the sky.

The errors of *Afρ* estimations were calculated using of *R* +σ and *R* −σ values (Equation 1) and averaged.

Table 1  Parameters of cometary orbits

| Comet | q (AU) | a (AU) | e | i (°) | P (yr) | $T_J$ |
|---|---|---|---|---|---|---|
| *Main Belt Comets* | | | | | | |
| 133P/Elst – Pizarro | 2.66 | 3.16 | 0.16 | 1.39 | 5.63 | 3.18 |
| 238P/Read | 2.37 | 3.17 | 0.25 | 1.26 | 5.63 | 3.15 |
| 288P/(300163) | 2.44 | 3.05 | 0.20 | 3.24 | 5.32 | 3.20 |
| P/2013 R3 (Catalina – PANSTARRS) | 2.20 | 3.03 | 0.27 | 0.87 | 5.28 | 3.18 |
| *Active Asteroids* | | | | | | |
| 311P/PANSTARRS | 1.94 | 2.19 | 0.12 | 4.97 | 3.24 | 3.66 |
| 331P/Gibbs | 2.89 | 3.01 | 0.04 | 9.74 | 5.22 | 3.23 |
| *Quasi-Hilda Comets* | | | | | | |
| 65P/Gunn | 2.91 | 3.88 | 0.25 | 9.19 | 7.64 | 2.99 |
| 74P/Smirnova – Chernyh | 3.54 | 4.16 | 0.15 | 6.65 | 8.48 | 3.01 |
| 117P/Helin – Roman – Alu 1 | 3.05 | 4.09 | 0.25 | 8.70 | 8.28 | 2.97 |
| 119P/Parker – Hartley | 2.84 | 4.19 | 0.32 | 5.57 | 8.59 | 2.93 |
| 246P/2010 V2 (NEAT) | 2.87 | 4.02 | 0.29 | 15.99 | 8.06 | 2.91 |
| 340P/Boattini | 3.06 | 4.25 | 0.28 | 2.08 | 8.76 | 2.96 |
| 345P/LINEAR | 3.16 | 4.05 | 0.22 | 2.72 | 8.14 | 3.00 |
| *Jupiter-Family Comets* | | | | | | |
| P/2014 E1 (Larson) | 2.14 | 3.71 | 0.42 | 15.98 | 7.16 | 2.87 |
| P/2014 MG4 (Spacewatch – PANSTARRS) | 3.71 | 5.01 | 0.26 | 9.37 | 11.2 | 2.91 |

The radii of the nuclei were estimated using the formula for estimating the size of planetary objects (Russel 1916):

$$R_n^2 = 2.238 \cdot 10^{22} r^2 \Delta^2 10^{0.4(m_\odot - m_{nucl} + \alpha\beta)} A^{-1}, \quad (2)$$

where r (AU) is the heliocentric distance; Δ (AU) is the geocentric distance; $m_{nucl}$ is the apparent *R* nuclear magnitude of the comet; α [deg] is the phase angle and β = 0.04 [mag/deg] the linear phase coefficient (Lamy et al. 2004). For the geometric albedo in the *R* filter we adopted the classical value of A = 0.04 for the quasi-Hilda comets (as for Jupiter family comets) and a value of A = 0.05 for the main-belt comets (as for C-type asteroids) (Hsieh et al. 2009a). The solar *R* magnitude used is $m_\odot$ = – 27.29 mag (Cox 2000). Samples of images are presented in Fig. 1.

Orbital parameters of the comets are listed in Table 1. A log of observations is listed in Table 2. We obtained color information only for several targets, mostly because of their low brightness and short time of visibility. The results of *BVR* photometry and color indices are shown in Table 3.

## 5  Results

### 5.1  Main Belt Comets

*133P/Elst – Pizarro* is the first main-belt object for which cometary activity was reported. This object was discovered in 1979, but its comet-like appearance was first detected in 1996 (Boehnhardt et al. 1996). The comet has an orbit lying within the asteroid belt and displayed a comet-like dust tail near perihelion in 1996, 2001, and 2007 (Hsieh et al. 2010; Jewitt et al. 2014b; Licandro et al. 2011). Our observations of 133P during its apparition in 2013 did not reveal any tail. However, the object occasionally displayed a coma of about 6″– 8″in diameter. Dust activity was lower than in previous apparitions (*Afρ* less than 10 cm in most cases).

*238P/Read (P/2005 U1)* has an orbit within the asteroid belt and displays a coma. It fits the definition of an Encke-type comet with ($T_J$ > 3; *a* < *aJ* ). Comet 238P was dis-covered on October 24, 2005, three months after perihelion passage, when it showed strong cometary activity extending until 2005 December 27. Usually, 238P is more active than 133P/Elst-Pizarro. Observations during the inactivity period (Hsieh et al. 2011) showed that it has a small nucleus of about 0.8 km in diameter. In 2016, we observed high cometary activity (high variability in color observed, Table 3) like during its 2005 apparition (Hsieh et al. 2009b). But *Afρ* was still low ( ≤ 3.2 cm).

*288P/(300163)*. This object showed a 6″-long dust tail in WFC3 images obtained with the Hubble Space Telescope on Aug. 22, 2016 (Agarwal et al. 2016). Comet 288P had been active during its previous perihelion passage in 2011 but returned to inactivity in 2012. The repeated appearance of a dust tail at a comparable orbital position in two consecutive orbits suggests that the activity is driven by the sublimation of ice, triggered by the enhanced solar heating near perihelion (Agarwal et al. 2016). The binary nature of 288P was discovered in December 2011 by J. Agarwal, D. Jewitt, M. Mutchler, H. Weaver, S. Larson using Hubble Space Telescope observations. Both components are similar in mass and size, making it a true binary system. The components are estimated to measure 1.8 kilometres in diameter, orbiting each other at a wide separation of 104 kilometres

Table 2  Log of *R* band observations and results of measurements

| Comet | Date | $N_i \times$ Expos. (s) | r (AU) | Δ (AU) | α(°) | R (mag) | Afρ (cm) | ρ('') | Observatory |
|---|---|---|---|---|---|---|---|---|---|
| 65P | 2012-10-14 | 8×300 | 4.60 | 3.91 | 9.7 | 19.93 ± 0.14 | 35 ± 5 | 3.8 | G36 |
|  | 2013-09-23 | 23×300 | 4.89 | 4.97 | 11.6 | 19.33 ± 0.08 | 44 ± 3 | 7.5 | G36 |
|  | 2013-11-15 | 26×300 | 4.90 | 4.20 | 8.9 | 19.25 ± 0.12 | 40 ± 4 | 7.5 | G36 |
|  | 2013-11-26 | 9×300 | 4.90 | 4.08 | 7.1 | 18.85 ± 0.12 | 56 ± 6 | 7.5 | G36 |
|  | 2014-03-06 | 6×300 | 4.88 | 4.54 | 11.3 | 19.12 ± 0.07 | 97 ± 6 | 3.8 | G36 |
|  | 2014-12-20 | 7×300 | 4.67 | 3.94 | 8.9 | 18.73 ± 0.09 | 55 ± 5 | 7.5 | G36 |
| 74P | 2013-09-23 | 10×300 | 4.78 | 3.86 | 5.4 | 18.85 ± 0.06 | 51 ± 3 | 7.5 | G36 |
|  | 2013-09-27 | 3×120 | 4.78 | 3.89 | 6.1 | 17.93 ± 0.12 | 118 ± 13 | 7.6 | 585 |
|  | 2014-08-25 | 8×300 | 4.69 | 3.96 | 9.3 | 18.99 ± 0.06 | 66 ± 4 | 5.0 | G36 |
|  | 2016-09-10 | 16×60 | 3.91 | 4.23 | 13.5 | 18.17 ± 0.08 | 92 ± 7 | 5.7 | 585 |
| 117P | 2013-04-02 | 8×60 | 3.49 | 2.59 | 8.2 | 16.81 ± 0.05 | 314 ± 14 | 2.8 | 585 |
|  | 2013-06-24 | 14×60 | 3.33 | 2.71 | 15.4 | 14.87 ± 0.04 | 446 ± 16 | 11.4 | 585 |
|  | 2013-08-10 | 7×60 | 3.25 | 3.26 | 17.9 | 15.08 ± 0.07 | 361 ± 23 | 13.3 | 585 |
| 119P | 2013-09-28 | 10×120 | 3.18 | 2.29 | 9.6 | 17.71 ± 0.11 | 30 ± 3 | 9.5 | 585 |
|  | 2013-10-31 | 9×120 | 3.13 | 2.15 | 2.2 | 17.05 ± 0.05 | 63 ± 3 | 7.6 | 585 |
|  | 2013-12-10 | 19×120 | 3.09 | 2.37 | 15.2 | 17.53 ± 0.06 | 44 ± 2 | 7.6 | 585 |
|  | 2013-12-24 | 13×120 | 3.07 | 2.53 | 16.9 | 17.54 ± 0.07 | 36 ± 2 | 9.5 | 585 |
|  | 2014-01-29 | 10×120 | 3.05 | 2.99 | 18.8 | 18.02 ± 0.10 | 34 ± 3 | 7.6 | 585 |
|  | 2014-02-03 | 12×120 | 3.04 | 3.06 | 18.6 | 18.26 ± 0.10 | 28 ± 2 | 7.6 | 585 |
| 133P | 2012-06-22 | 11×300 | 2.81 | 2.05 | 16.1 | 20.23 ± 0.07 | 2.6 ± 0.2 | 7.5 | G36 |
|  | 2012-07-09 | 8×300 | 2.79 | 2.22 | 19.4 | 20.80 ± 0.09 | 1.6 ± 0.1 | 7.5 | G36 |
|  | 2013-09-10 | 2×300 | 2.79 | 1.79 | 2.8 | 19.29 ± 0.07 | 5.3 ± 0.3 | 7.5 | G36 |
|  | 2013-09-28 | 4×300 | 2.82 | 1.89 | 9.7 | 19.66 ± 0.10 | 4.1 ± 0.4 | 7.5 | G36 |
|  | 2013-11-06 | 4×300 | 2.87 | 2.34 | 18.6 | 21.26 ± 0.09 | 1.2 ± 0.1 | 7.5 | G36 |
|  | 2013-11-15 | 11×300 | 2.88 | 2.47 | 19.4 | 20.53 ± 0.14 | 2.5 ± 0.3 | 7.5 | G36 |
|  | 2014-10-30 | 48×60 | 3.40 | 2.51 | 8.6 | 20.01 ± 0.17 | 11 ± 2 | 3.8 | 585 |
| 238P | 2016-10-28 | 5×300 | 2.37 | 1.45 | 11.7 | 19.53 ± 0.11 | 2.5 ± 0.3 | 7.5 | G36 |
|  | 2016-11-01 | 4×300 | 2.37 | 1.47 | 13.3 | 19.29 ± 0.10 | 3.2 ± 0.3 | 7.5 | G36 |
| 246P | 2012-02-22 | 17×300 | 3.37 | 2.64 | 12.7 | 14.79 ± 0.03 | 2169 ± 60 | 2.5 | G36 |
|  | 2012-03-24 | 8×300 | 3.30 | 2.36 | 6.8 | 14.20 ± 0.03 | 3201 ± 88 | 2.5 | G36 |
|  | 2014-08-29 | 2×300 | 3.96 | 3.10 | 8.7 | 16.19 ± 0.06 | 969 ± 54 | 2.5 | G36 |
| 288P | 2016-10-31 | 12×300 | 2.44 | 1.74 | 19.9 | 19.16 ± 0.05 | 4.5 ± 0.2 | 7.5 | G36 |
|  | 2016-11-01 | 2×300 | 2.44 | 1.75 | 20.1 | 19.25 ± 0.10 | 4.1 ± 0.4 | 7.5 | G36 |
|  | 2016-12-26 | 4×300 | 2.45 | 2.43 | 23.3 | 19.81 ± 0.07 | 3.5 ± 0.2 | 7.5 | G36 |
| 311P | 2013-09-11 | 11×300 | 2.11 | 1.12 | 5.5 | 19.96 ± 0.05 | 3.1 ± 0.1 | 2.5 | G36 |
|  | 2013-09-25 | 7×300 | 2.09 | 1.15 | 12.2 | 19.57 ± 0.06 | 1.5 ± 0.1 | 7.5 | G36 |
|  | 2013-10-07 | 6×300 | 2.08 | 1.20 | 17.7 | 20.16 ± 0.10 | 0.9 ± 0.1 | 7.5 | G36 |
| 331P | 2012-04-22 | 6×300 | 3.11 | 2.33 | 13.5 | 20.26 ± 0.11 | 3.5 ± 0.4 | 7.5 | G36 |
|  | 2012-06-23 | 2×300 | 3.12 | 3.13 | 18.7 | 21.27 ± 0.20 | 1.9 ± 0.4 | 7.5 | G36 |
| 340P | 2016-09-07 | 20×60 | 3.09 | 2.30 | 13.7 | 18.73 ± 0.11 | 18 ± 2 | 5.7 | 585 |
|  | 2016-10-30 | 5×300 | 3.07 | 2.95 | 18.9 | 18.93 ± 0.07 | 22 ± 1 | 5.0 | G36 |
| 345P | 2016-09-07 | 14×120 | 3.17 | 2.22 | 7.2 | 18.50 ± 0.10 | 18 ± 2 | 7.6 | 585 |
|  | 2016-10-28 | 8×300 | 3.19 | 2.33 | 10.2 | 19.48 ± 0.13 | 15 ± 2 | 3.8 | G36 |
|  | 2016-11-01 | 12×300 | 3.20 | 2.36 | 11.3 | 18.88 ± 0.09 | 13 ± 1 | 7.5 | G36 |
|  | 2016-12-27 | 4×300 | 3.24 | 3.12 | 17.6 | 19.74 ± 0.11 | 8.0 ± 0.8 | 7.5 | G36 |
| P/2013 R3 | 2013-10-31 | 10×120 | 2.27 | 1.36 | 18.9 | 17.77 ± 0.08 | 8.6 ± 0.6 | 9.5 | 585 |
|  | 2013-11-17 | 9×300 | 2.29 | 1.52 | 19.0 | 18.51 ± 0.09 | 6.2 ± 0.5 | 7.5 | G36 |
|  | 2013-12-24 | 12×120 | 2.36 | 1.98 | 24.3 | 18.95 ± 0.12 | 5.7 ± 0.6 | 7.6 | 585 |
| P/2014 E1 | 2014-03-13 | 5×120 | 2.21 | 1.41 | 19.4 | 17.87 ± 0.09 | 9.7 ± 0.8 | 7.6 | 585 |
|  | 2014-03-22 | 20×60 | 2.19 | 1.32 | 16.6 | 16.62 ± 0.08 | 23 ± 2 | 9.5 | 585 |
|  | 2014-04-23 | 10×120 | 2.15 | 1.15 | 1.4 | 16.59 ± 0.13 | 19 ± 2 | 9.5 | 585 |
|  | 2014-04-26 | 12×60 | 2.15 | 1.14 | 1.0 | 16.07 ± 0.05 | 31 ± 1 | 9.5 | 585 |
| P/2014 MG4 | 2014-08-18 | 11×120 | 4.16 | 3.19 | 4.2 | 18.13 ± 0.10 | 81 ± 8 | 5.7 | 585 |

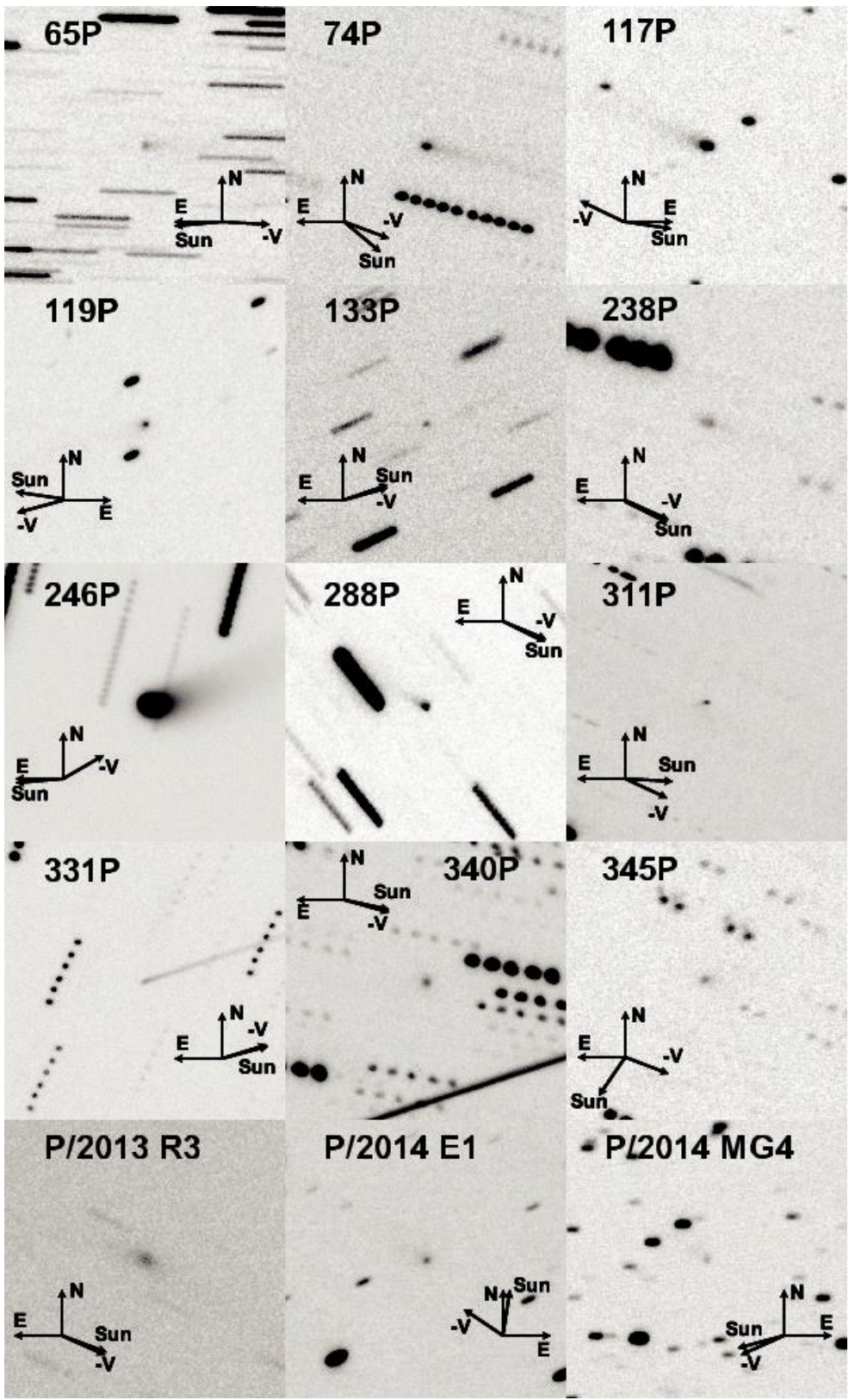

Fig. 1 R-band images of 15 observed comets. The FOV is $2.1' \times 2.1'$ for Calar Alto Observatory (G36) and $3.2' \times 3.2'$ for Kyiv Comet Station (585). The observing dates and geometry are shown in Table 2.

Table 3 Log of multiband observations and results of color measurements

| Date | r (AU) | Δ (AU) | α(°) | B (mag) | V (mag) | R (mag) | B–V | V–R | ρ(″) | Observ. |
|---|---|---|---|---|---|---|---|---|---|---|
| 119P/Parker – Hartley ||||||||||| 
| 2014-02-03 | 3.04 | 3.06 | 18.6 | — | 18.61±0.12 | 18.26±0.10 | — | 0.35±0.16 | 7.6 | 585 |
| 238P/Read ||||||||||| 
| 2016-10-28 | 2.37 | 1.45 | 11.7 | 20.74±0.25 | 19.94±0.15 | 19.53±0.11 | 0.80±0.29 | 0.41±0.19 | 7.5 | G36 |
| 2016-11-01 | 2.37 | 1.47 | 13.3 | 20.44±0.26 | 19.97±0.18 | 19.29±0.10 | 0.47±0.32 | 0.68±0.21 | 7.5 | G36 |
| 288P/(300163) ||||||||||| 
| 2016-10-31 | 2.44 | 1.74 | 19.9 | 20.30±0.26 | 19.61±0.11 | 19.16±0.05 | 0.69±0.28 | 0.45±0.12 | 7.5 | G36 |
| 2016-11-01 | 2.44 | 1.75 | 20.1 | 20.31±0.26 | 19.54±0.13 | 19.25±0.10 | 0.77±0.29 | 0.29±0.16 | 7.5 | G36 |
| 2016-12-26 | 2.45 | 2.43 | 23.3 | 20.96±0.25 | 20.16±0.11 | 19.81±0.07 | 0.80±0.27 | 0.35±0.13 | 7.5 | G36 |
| 340P/Boattini ||||||||||| 
| 2016-09-07 | 3.09 | 2.30 | 13.7 | — | 19.11±0.14 | 18.73±0.11 |  | 0.38±0.18 |  | 585 |
| 2016-10-30 | 3.07 | 2.95 | 18.9 | 20.18±0.15 | 19.46±0.07 | 18.93±0.07 | 0.72±0.17 | 0.53±0.10 | 5.0 | G36 |
| 2016-10-31 | 3.06 | 2.96 | 18.9 | 20.09±0.14 | 19.26±0.09 | — | 0.83±0.17 | — | 5.0 | G36 |
| 345P/LINEAR) ||||||||||| 
| 2016-10-28 | 3.19 | 2.33 | 10.2 | 20.95±0.24 | 20.07±0.23 | 19.48±0.13 | 0.88±0.33 | 0.59±0.26 | 3.8 | G36 |
| 2016-11-01 | 3.20 | 2.36 | 11.3 | 20.32±0.28 | 19.31±0.22 | 18.88±0.09 | 1.01±0.36 | 0.43±0.24 | 7.5 | G36 |
| 2016-12-27 | 3.24 | 3.12 | 17.6 | 21.23±0.23 | 20.21±0.16 | 19.74±0.11 | 1.02±0.28 | 0.47±0.19 | 7.5 | G36 |

every 135 days (Agarwal et al. 2017). In December 2016, comet 288P had a tail about 16″ long as seen from our observations from Calar Alto. Estimations of $Af\rho$ were less than 5 cm, measured $B-V$ values were about $0.7-0.8$.

*P/2013 R3 (Catalina – PANSTARRS)*. The orbital parameters of this object classify it as a member of the main asteroid belt, although its fuzzy appearance resembles that of a comet. Its Tisserand parameter relative to Jupiter, $T_J = 3.18$, is larger than the nominal ($T_J = 3$) dividing line separating dynamical comets ($T_J < 3$) from asteroids ($T_J > 3$). In our 2013 observations, P/2013 R3 appeared as a diffuse object with a strong condensation and $Af\rho$ less than 10 cm. During the 2013 – 2014 apparition, the comet disintegrated into three fragments (Licandro et al. 2013). About 10 or more components were detected by the Hubble Space Telescope (Jewitt et al. 2014a).

### 5.2 Active asteroids

*311P/PANSTARRS* is an active asteroid discovered on August 27, 2013. Observations made by the Hubble Space Telescope revealed that it had six comet-like tails (Jewitt et al. 2013). The tails are suspected to be streams of surface material centrifugally ejected by the fast-spinning asteroid (Hainaut et al. 2014). Three-dimensional models show that the tails could have been formed by a series of periodic impulsive dust-ejection events, eventually stretched into streams by solar radiation pressure (Jewitt et al. 2015a). Precovery images from the Sloan Digital Sky Survey from 2005 showed negligible cometary activity in 2005. By Calar Alto observations, it was a compact object with very low $Af\rho$ (up to 0.9 cm in October 2013).

*331P/Gibbs*, similarly to 311P, showed the presence of multiple, 200 m-sized fragments during observations taken in 2014 (Drahus et al. 2015). Again, this phenomenon is interpreted as the ejection of material from its rapidly-spinning nucleus (P=3.24±0.01 h). In our 2012 observations, the object showed a narrow dust tail about 25″ long from our observations taken from Calar Alto. Dust activity was low ($Af\rho$ less than 5 cm).

### 5.3 Quasi-Hilda and Jupiter-Family Comets

*65P/Gunn* is a QHC, orbiting the Sun every 6.79 years between the orbits of the planets Mars and Jupiter. The Wide-field Infrared Survey Explorer (WISE) observed the comet on April 24, 2010, just one month after the comet's closest approach to the Sun. The WISE infrared images showed a leading and a trailing dust trail, which was interpreted as material shed from the surface of the object, and which rep-resents the first stage in the formation of a meteor stream (Walker et al. 2011). It was observed at Calar Alto observatory during 2012 – 2014 near aphelion, therefore, dust activity was relatively low ($Af\rho$ less than 100 cm). Other results (Lowry & Fitzsimmons 2001), obtained in previous apparitions when it was closer to the perihelion, suggest that the $Af\rho$ could be higher than our estimates.

*74P/Smirnova – Chernykh* is one of the brightest QHC with ($T_J > 3$; $a < a_J$). Our Calar Alto observations from 2013 showed that the comet had a tail of about 32″ and no coma. Observations of the comet during 2013 – 2016 showed relatively low dust activity ($Af\rho$ less than 120 cm).

*117P/Helin – Roman – Alu 1* was discovered in 1989, two years after its perihelion passage in Octo-ber 1987. This QHC tends to increase its brightness in an extended period after the perihelion passage (http://www.aerith.net/comet/catalog/0117P/index.html). During the 1997 apparition, it reached maximum brightness nearly one year after perihelion.

Kazuo Kinoshita's calculations (http://jcometobs.web.fc2.com/pcmtn/0117p.htm) reveal that the comet passed 0.68 AU from Jupiter in 2002, and the perihelion distance was reduced from 3.7 AU to 3.0 AU. In the following return in 2005, it became brightest about 100 days after the perihelion passage. The difference between the perihelion passage and the brightest day was reduced because the comet approached closer to the Sun. Pre-perihelion observations at Kyiv Comet Station in 2013 showed $Af\rho$ values higher than 300 cm.

*119P/Parker – Hartley* is a QHC with a relatively circular orbit. It was observable in 2013 – 2014 as a condensed object. During our observations in October 2013, the comet had a tail of 30″. We obtained $Af\rho$ values about 30 – 60 cm.

*246P/NEAT* is a QHC, which, due to its high brightness, is observable throughout its orbit, excluding solar conjunction. We observed it in spring 2012 while it was a bright object with long dust tail up to 95″. The comet demonstrated extremely high dust activity (sometimes $Af\rho$ value was more than 3000 cm at pre-perihelion.)

*340P/Boattini* is a QHC with low orbit inclination ($i = 2.08$). We observed it in September – October 2016 as an asteroidal object with a tiny coma and no tail. Measured $B - V$ values were about 0.7 – 0.8.

*345P/LINEAR* is a QHC discovered during the Mt. Lemmon Survey with the 1.5-m reflector in August 2016 was linked to asteroid 2008 SH164 discovered by LINEAR in September 2008. It was observed in September – December 2016 as a faint asteroidal object with a compact coma with-out a tail. Our Calar Alto multiband observations showed some high $B - V$ values (about 0.9 – 1.0) (Table 3).

*P/2014 E1 (Larson)* is a JFC with a perihelion distance of 2.14 AU. Although not strictly a QHC, its orbit lies for a good fraction within the main asteroid belt and therefore was included in our observation program. This comet was active during its discovery apparition. During our observations in April 2014, it showed bright condensation with 20″ coma and about 70″ tail. $Af\rho$ values were slowly growing from 9.7 cm to 31 cm at pre-perihelion.

*P/2014 MG4 (Spacewatch – PANSTARRS)* is a JFC dis-covered in 2014, one year after the perihelion passage. The comet was active in 2014 – 2015 with an outburst brightening of about 3 mag. During our observations in August 2014, it had a condensed 7″ coma and $Af\rho = 81$ cm.

## 6 Discussion

### 6.1 $Af\rho$

We calculated $Af\rho$ values for 15 comets in the main belt and surrounding regions. As a rule, increasing levels of cometary activity lead to increasing $Af\rho$ values. The values we measured for MBAOs were rather low – less than 10 cm (except outbursts, like for 133P) (Table 2). On the other hand, QHCs and JFCs displayed higher levels of $Af\rho$ – from several dozens to several hundred centimetres – which points to possible different activity mechanisms as for MBAOs.

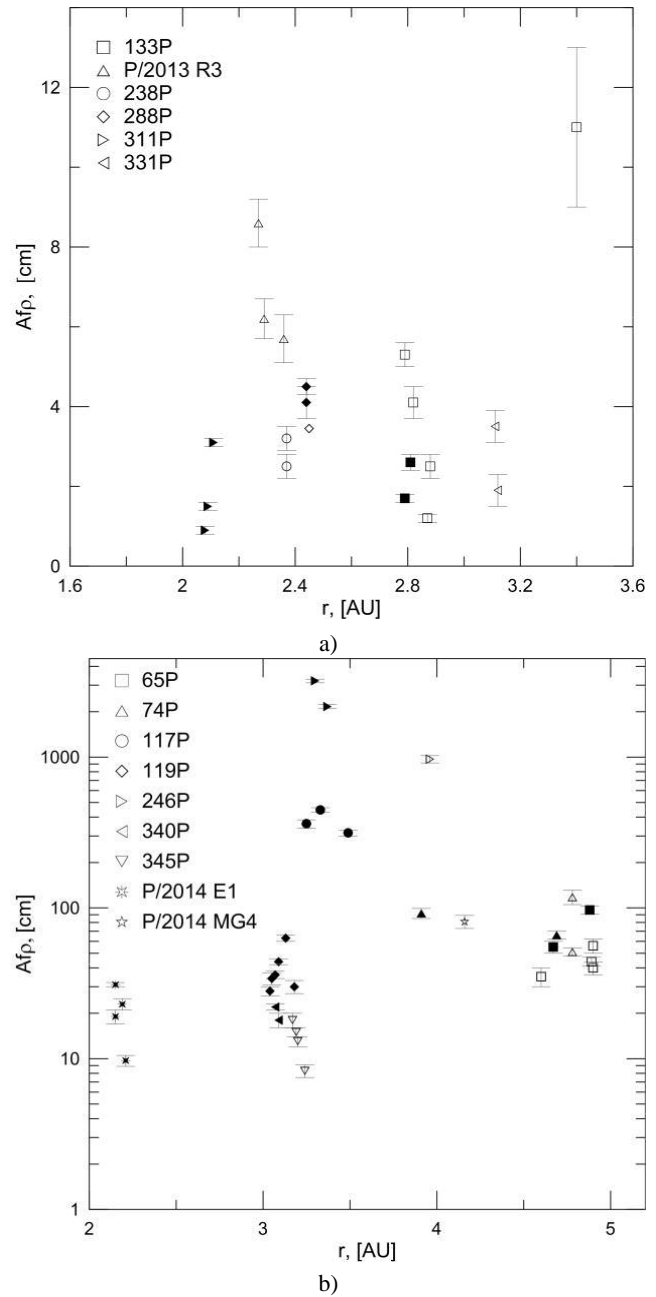

Fig. 2 Heliocentric distribution of $Af\rho$ values for selected comets: a)MBCs and AAs, b)QHCs and JFCs. Filled and open symbols represent pre- and post-perihelion data, respectively.

Other mechanisms could potentially produce a comet-like mass loss on AAs and MBCs, including thermal fracturing, radiation pressure sweeping, and electrostatic levitation. Furthermore, it is important to consider that activity for the given bodies could be produced by a combination of effects, and individual active episodes on the same active asteroid or main-belt comet could even have different sets of driving mechanisms each time (Jewitt et al. 2015c; Hsieh 2017).

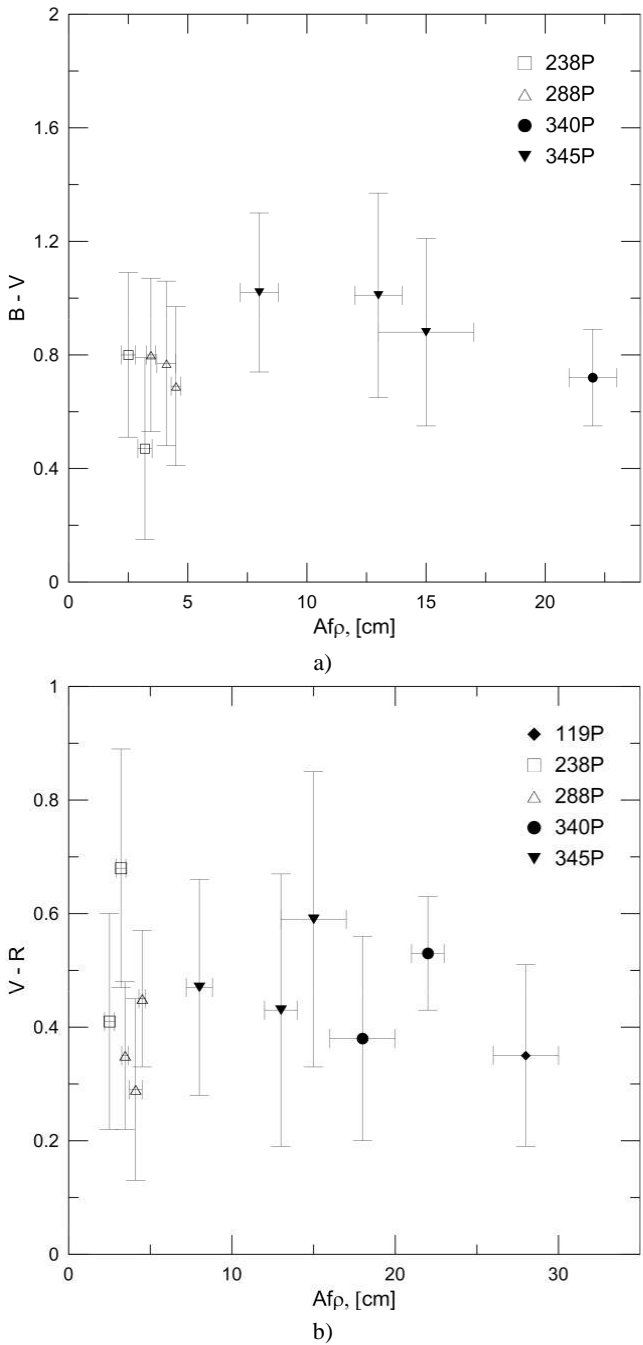

Fig. 3 Color index $B - V$ (a) and $V - R$ (b) over dust activity $Af\rho$. Filled symbols represent QHCs.

An important factor to keep in mind when comparing $Af\rho$ results is that the heliocentric distance for QHCs is about twice as large as that for MBCs. Also, QHCs tend to be larger than MBCs (Fernandez et al. 2013; Hsieh 2014). Fig. 2 shows the distribution of $Af\rho$ as a function of heliocentric distance for MBCs, AAs, QHCs, and JFCs. Continuous decrease of $Af\rho$ was observed for 345P, P/2013 R3, and sometimes for 133P at post-perihelion. For some objects, we have close values of $Af\rho$ for the pre- and post-perihelion positions (133P, 65P, 74P). Pre- or post-perihelion effects of anomalously high dust activity are not observed in either group (Fig. 2). However, the available observational data are too scarce to determine any significant differences between pre- and post-perihelion behaviour of the comets.

### 6.2 Color indices

Color indices as a function of $Af\rho$ for the observed objects are shown in Fig. 3. Most of $V - R$ color estimations for QHCs lay in the range of about 0.4 – 0.6, in agreement with typical JFCs (Borysenko et al. 2019; Jewitt 2015). Color indices of comets may depend on the dust activity of their atmospheres. On the other hand, cometary colors depend on the properties of dust particles of the cometary atmospheres (Jewitt 2015). Large color estimation errors do not allow making an obvious statement about the reddening effect for observed comets. However, comet 345P/LINEAR observed at a maximum of activity showed $B - V$ indices about 0.9 – 1.0.

### 6.3 Sizes of MBCs, AAs and QHCs

We estimated the radii of the cometary nuclei listed in Table 4 by applying the coma compensation method described in Hicks et al. (2007) and references therein. In this method, a stellar PSF obtained from background stars is used to fit the innermost coma. Whenever observations at multiple epochs were available, we only used those observations with the lowest activity level. In the case of faint activity, the resulting radius is a good estimation of the nuclear radius, whereas, in the case when coma dominated in observations, the resulting radius can be considered an upper bound only.

For the AAs in our sample, the radii determinations range from 0.4 km for 311P to 1.7 km for 331P and are somewhat larger than the previous determinations by Jewitt et al. (2013) and Drahus et al. (2015).

Our estimations for the radii for 65P and 74P improve the upper limits by Lowry & Fitzsimmons (2001) and are closer to the determinations by Tancredi et al. (2000) and Lamy et al. (2004), respectively. For 117P, 119P, and 246P, we could only obtain upper bounds of radii, due to considerable activity at the time of the observations. In any case, our upper limits are compatible with the measurements by Tancredi et al. (2000) and Fernandez et al. (2013) (see Table 4).

We did not estimate the radii of the cometary nuclei for some objects – comet P/2013 R3 (Catalina – PANSTARRS) was splitting during our observations, and 288P/(300163) is a binary object which couldn't be resolved by our telescope.

Table 4 Physical parameters of cometary nuclei

| Comet | Type | Date | $m_{nucl}$ (mag) | $R_n$ (km) | Prev. $R_n$ | Reference |
|---|---|---|---|---|---|---|
| 65P/Gunn | QHC | 2012-10-14 | 21.53 ± 0.06 | 2.8 ± 0.1 | 4.8 | Tancredi et al. (2000) |
| 74P/Smirnova-Chernyh | QHC | 2014-08-25 | 21.16 ± 0.05 | 3.4 ± 0.1 | 2.2 | Lamy et al. (2004) |
| 117P/Helin-Roman-Alu 1 | QHC | 2013-04-02 | 16.47 ± 0.05 | < 13.9 ± 0.3 | 3.5 | Tancredi et al. (2000) |
| 119P/Parker-Hartley | QHC | 2014-02-03 | 19.24 ± 0.08 | < 4.8 ± 0.2 | 0.98 | Fernandez et al. (2013) |
| 133P/Elst-Pizarro | MBC | 2013-11-06 | 21.26 ± 0.09 | 1.2 ± 0.1 | 1.6 | Bauer et al. (2012) |
| 238P/Read | MBC | 2016-10-28 | 19.53 ± 0.07 | 1.4 ± 0.1 | 0.4 | Hsieh et al. (2011) |
| 246P/2010 V2 (NEAT) | QHC | 2014-08-29 | 17.81 ± 0.04 | < 10.3 ± 0.2 | 4.2 | Fernandez et al. (2013) |
| 311P/PANSTARRS | AA | 2013-10-07 | 21.48 ± 0.07 | 0.4 ± 0.1 | < 0.24 | Jewitt et al. (2013) |
| 331P/Gibbs | AA | 2012-06-23 | 21.43 ± 0.18 | 1.7 ± 0.2 | 0.9 | Drahus et al. (2015) |
| 340P/Boattini | QHC | 2016-10-30 | 21.02 ± 0.04 | 2.1 ± 0.1 | – | – |
| 345P/LINEAR | QHC | 2016-09-07 | 19.20 ± 0.08 | 3.0 ± 0.1 | – | – |
| P/2014 E1 (Larson) | JFC | 2014-04-26 | 17.36 ± 0.05 | 2.2 ± 0.1 | – | – |
| P/2014 MG4 (Spacewatch-PANSTARRS) | JFC | 2014-08-18 | 18.32 ± 0.07 | < 5.5 ± 0.2 | – | – |

## 7 Conclusions

In the period 2012 – 2016, we obtained $Af\rho$ values for 15 active objects residing in or in the vicinity of the main belt. We also obtained estimations of radii for 13 of those objects. Our observations show systematic differences in the measured parameters for two classes of objects. $Af\rho$ values for QHCs are in the range of those of other Jupiter family comets (from a few dozens to several hundred centimetres; Lowry et al. 1999; Lowry & Fitzsimmons 2001; Lowry et al. 2003; Borysenko et al. 2019). On the other hand, MBCs and AAs show smaller values of $Af\rho$ (usually less than 10 cm). Also, our measurements seem to confirm that the sizes of AA and MBC nuclei are smaller than those of QHCs and JFCs.

In the period October-December, 2016, BVR images were obtained for 4 MBCs and QHCs. The resulting $B - V$ colors range from about 0.5 to 1.0 (Table 3). However, the accuracy and the size of this sample is not sufficient to address any possible systematic differences between MBCs and QHCs.

*Acknowledgements.* Based on observations obtained at the Kyiv Comet Station and the Calar Alto Observatory. The German-Spanish Astronomical Center at Calar Alto is operated jointly by the Max-Planck-Institut für Astronomie (MPIA) in Heidelberg, Germany, and the Instituto de Astrofisica de Andalucia (CSIC) in Granada, Spain. This work was supported by the Deutscher Akademischer Austauschdienst (DAAD) grant. Karen Jean Meech acknowledges support through National Science Foundation grant AST1617015.